\documentclass[11pt]{article}
\usepackage{xspace}
\usepackage{graphicx}
\usepackage{amsmath}
\usepackage{amssymb}
\usepackage{color}

\definecolor{Red}{rgb}{1,0,0}
\definecolor{Green}{rgb}{0,1,0}
\definecolor{Blue}{rgb}{0,0,1}
\definecolor{Black}{rgb}{0,0,0}

\def\beq{\begin{equation}}
\def\eeq#1{\label{#1}\end{equation}}
\def\eeqn{\end{equation}}

\def\beqa{\begin{eqnarray}}
\def\eeqa#1{\label{#1}\end{eqnarray}}
\def\eeqan{\end{eqnarray}}

\let\bar=\overbar

\def\Dslash{\not{\hbox{\kern-4pt $D$}}}
\def\dslash{\not{\hbox{\kern-2pt $\del$}}}

\def\msb{{\bar{\ssstyle M \kern -1pt S}}}

\def\Title#1{\begin{center} {\Large {\bf #1} } \end{center}}

\begin{document}

\Title{Indirect Detection of WIMP Dark Matter: a compact review}

\bigskip\bigskip


\begin{raggedright}  

Jan Conrad \index{Conrad, J.}, {\it Oskar Klein Centre, Physics Department, Stockholm University, Albanova, SE-10691 Stockholm,  Sweden}\\

\bigskip
\end{raggedright}

\begin{abstract}
Indirect detection of dark matter particles, i.e. the detection of annihilation or decay products of Weakly Interacting Massive Particles, has entered a pivotal phase as experiments reach sensitivities that probe the most interesting parameter space. This period is naturally accompanied by claims of detection. In this contribution I discuss and compare different probes (gamma-rays, neutrinos and charged cosmic rays) and review the status and prospects of constraints and recent detection claims.

\end{abstract}

{\small
\begin{flushleft}
\emph{To appear in the proceedings of the Interplay between Particle and Astroparticle Physics workshop, 18 -- 22 August, 2014, held at Queen Mary University of London, UK.}
\end{flushleft}
}

\section{Introduction}
Cosmological observations now proof beyond reasonable doubts that around 85 \% of the matter component of the Universe is comprised of new type of matter, dubbed dark matter (DM), see e.g. \cite{Bergstrom:2012fi}, consisting of particle(s) not currently part of the standard model. The currently most popular, almost paradigmatic, candidate is a weakly interacting massive particle (WIMP), i.e. a particle with weak interactions and masses roughly above the mass of the proton. The reason for this paradigmatic status of the WIMP is that thermal WIMP production in the big bang, whose processes are well gauged by the observations of light elements, predict a global DM abundance within one dex of the observed one (e.g. \cite{Jungman:1995df}). A result often called the ``WIMP miracle''. There are three ways to try to find WIMPs. Attempts to produce WIMPs are undertaken at the Large Hadron Collider, especially anticipating results of the new data at 14 TeV center of mass energy. WIMPs scattering of deep underground low background detectors (e.g. \cite{Baudis:2013eba}) is dubbed \emph{direct detection}. Finally, the approach discussed here, is to observe that WIMPs might annihilate (or decay)  in dense region of the Universe to yield standard model particles, in particular gamma rays, charged leptons and neutrinos. For gamma rays and neutrinos, that essentially travel through space undisturbed, the resulting flux is given by:
\begin{equation}
\frac{dR}{dt\,dA \,dE} =  P  \cdot J (\Delta \Omega)
\end{equation}
with $R$ being the number of particles and $P$ and $J$ defined as:
\begin{equation}
P = \frac{\left<\sigma_{ann}\rm{v}\right>}{2m_{\chi}^2} \cdot \sum_i BR_i \frac{dN^i_\gamma}{dE_i} 
\label{eq:source_indirect}
\end{equation}
with  $(\sigma_{ann}\rm{v})$ being the annihilation cross-section averaged over the velocity distribution of WIMPs, $m_{\chi}$ is the WIMP  mass, $BR_i$ denotes the  branching fraction to different annihilation channels (e.g. quarks and anti-quarks) and  $\frac{dN^i_\gamma}{dE_i}$ is the yield of particles as function of energy, determining the spectral shape of the signal. Different models for WIMPs imply different branching fractions to different final state particles. For quarks the energy spectrum is relatively independent of flavor \cite{Fornengo:2004kj}, which is why usually branching fractions of one into a given quark state (commonly b-quark pairs), and separately for lepton states, are used to calculate constraints in the cross-section vs. WIMP mass plane. $P$ is the particle physics factor, i.e. contains the particle properties affecting the potential signal.  The other term represents the respective astrophysical part:
\begin{equation}
J(\Delta \Omega) = \int_{\Delta \Omega}\int_{l=0}^{\infty} dl \,d\Omega  \rho_\chi^2 (l) 
\label{eq:J-factor}
\end{equation}
with $\rho_\chi$ being the DM density along the line of sight, $\Delta \Omega$ the solid angle element that is integrated over in the observation. Charged leptons diffuse through the Galactic medium suffering various energy loss mechanisms and interaction, thus the flux has to be calculated solving a diffusion equation with WIMP production as one of the source terms.

\section{$\gamma$-ray probe: the golden channel}

\noindent
Gamma rays are produced by WIMP annihilation in a variety of ways: annihilation into quarks and gauge bosons with subsequent hadronization and  pion decay yields a continuum spectrum. Direct annihilation to gamma rays and virtual internal bremsstrahlung yield spectral features, which (escpially in case of the line) constitute a smoking-gun signal, essentially impossible to explain by other means than dark matter. These signatures are to be compared to the common power-law features of conventional astrophysical gamma-ray sources, or harder, sources with exponential cut-offs such as pulsars. 
Spatial information can be used to discriminate in diffuse emission. For example the more spherical emission that the Galactic DM  halo would exhibit as compared to conventional Galactic diffuse emission, or in using  information in the angular power spectrum in the isotropic gamma-ray background, exploiting the square density dependence of the DM signal as compared to (modulo bias) linear density dependence of the conventional gamma-ray flux, e.g. \cite{Ando:2005xg,Cuoco:2010jb}. For a comprehensive review, see \cite{Bringmann:2012ez}.\\

\noindent
In the energy range between about $\sim$ 100 MeV and several 100 GeV gamma rays are observed by pair-conversion telescopes on satellites, foremost the Fermi Large Area Telescope \cite{Atwood:2009ez}. Above 100 GeV, Imaging Air Cherenkov Telescopes (IACT), such as HESS\cite{Aharonian:2006pe}, MAGIC \cite{Aleksic:2011bx} and VERITAS \cite{Holder:2008ux}, become more sensitive. Whereas satellites are small with typical effective areas of a square meter, IACTs, utilizing Earth's atmosphere, have effective areas of the order of 10000 square meters. However, the latter are penalized by small field of view ($\sim$ 5 deg as compared to 2.4 sr for the Fermi-LAT) and small duty cycle. Typically IACTs observer for about 1000 hours a year and only a fraction of (at most) 10 \% of these observations are dedicated to DM. Fermi-LAT is also essentially background free, in contrast to IACTs which even after event selection generically is dominated by charged cosmic ray background.\\

\begin{figure}[!ht]
\begin{center}
\includegraphics[width=0.8\columnwidth]{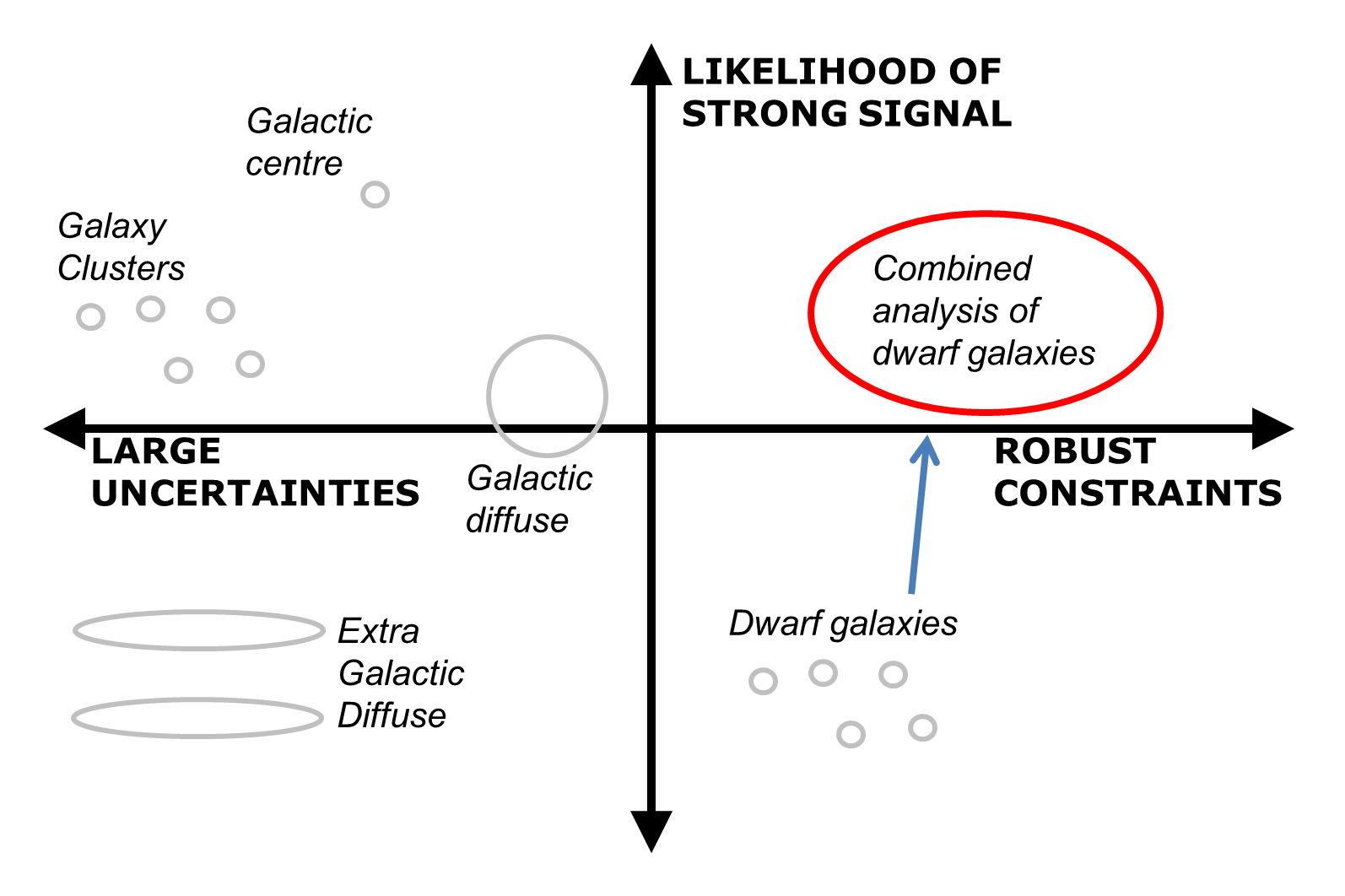}
\caption{An attempt to visualize the usefulness of different targets for gamma-ray (and neutrino) detection of WIMPs. A plane is defined of likelihood of strong signal versus robustness of constraints. Circles illustrate region of interest for analyses.}
\label{fig:targets}
\end{center}
\end{figure}

\noindent
As annihilation is proportional to DM density squared, density enhancements provide targets. In Fig. \ref{fig:targets} an attempt is made to illustrate the relative benefits and drawbacks of different targets by qualitatively distributing them on a plane of the likelihood to detect a strong signal from WIMPs versus the robustness of constraints if no signal is detected. We will begin by discussing Fermi-LAT searches and then move on to IACT searches.\\

\noindent
The robustness of constraints that can be obtained hinges foremost on the knowledge of the DM density and/or uncertainties in background modeling. For clusters of galaxies the expected flux is largely dependent on how substructure is modeled. N-body simulations can only resolve DM substructure to relatively large scales and extrapolations to smaller scales are necessary. Constraints on WIMP properties are therefore uncertain by orders of magnitude.  Similar arguments apply to a signal in the isotropic gamma-ray background. As gamma-ray sources increasingly get resolved the conventional background drops, but predictions for WIMP induced signal depend on modeling of substructure and their development via red shift as well as on assumptions on absorption \cite{Abdo:2010dk}. Background modeling is also the main problem for searches which try to use galactic diffuse emission \cite{Ackermann:2012rg}.  The Galactic Center (GC), possibly exhibiting a cusp in DM density, is expected to be providing the largest signal from WIMP annihilation. However, the uncertainty in the DM density in the innermost region, and more importantly the complexity of potential conventional gamma-ray sources in the region makes inference on WIMP properties subject to large uncertainties. This is less so for searches for spectral features, where the aim is to search for a deviation from a background expectation which can be determined data-driven, i.e. by side-band measurements, first done on Fermi data in \cite{Abdo:2010nc}. The claim of evidence for a line-line feature at about 130 GeV in 2012 \cite{Bringmann:2012vr,Weniger:2012tx,Su:2012ft} received considerable attention in the last years, even in part motivating a change in survey strategy of the Fermi-LAT. The most recent Fermi-LAT analyses confirm the presence of an excess however at much weaker significance \cite{Ackermann:2013uma, Albert:2014}, in addition the significance of the feature has been falling since the first publication, i.e. by the time of this conference, the physical origin of this feature seems doubtful. An ongoing independent analysis by HESSII, with its reduced energy threshold, should allow a final conclusion  \cite{Bergstrom:2012vd} within the next year. Despite the complexity of the GC region, the presence of an extended excess over the most commonly assumed diffuse emission model backgrounds and known sources at around 3 GeV  has led to a more recent claim of a tentative  WIMP detection \cite{Hooper:2010mq,Abazajian:2014fta,Daylan:2014rsa,Calore:2014xka}.  Dwarf galaxies, close-by companions of the Milky Way, which are DM dominated, on the other hand, provide robust target: there is in principle no astrophysical background and the WIMP density can relatively well be determined from measurements of stellar velocities. In contrast to clusters of galaxies, substructure is believed to be negligible, thus expected absolute fluxes are low. The realization that statistical methods can be developed to effectively combined dwarf galaxies \cite{Ackermann:2011wa,Ackermann:2013yva}, however, lead to dwarf galaxies providing the currently most stringent and robust constraints on WIMPs from indirect detection. The most recent result, based on the PASS8 event selection has been presented at the Fermi symposium in 2014 \cite{Anderson:2014}. The result provides exclusion of generic WIMPs (in the b-channel) up to masses of 100 GeV. It also reaches a sensitivity which could have provided confirmation of the DM interpretation of the above mentioned GC excess, but it did not.

\begin{figure}[!ht]
\begin{center}
\includegraphics[width=0.49\columnwidth]{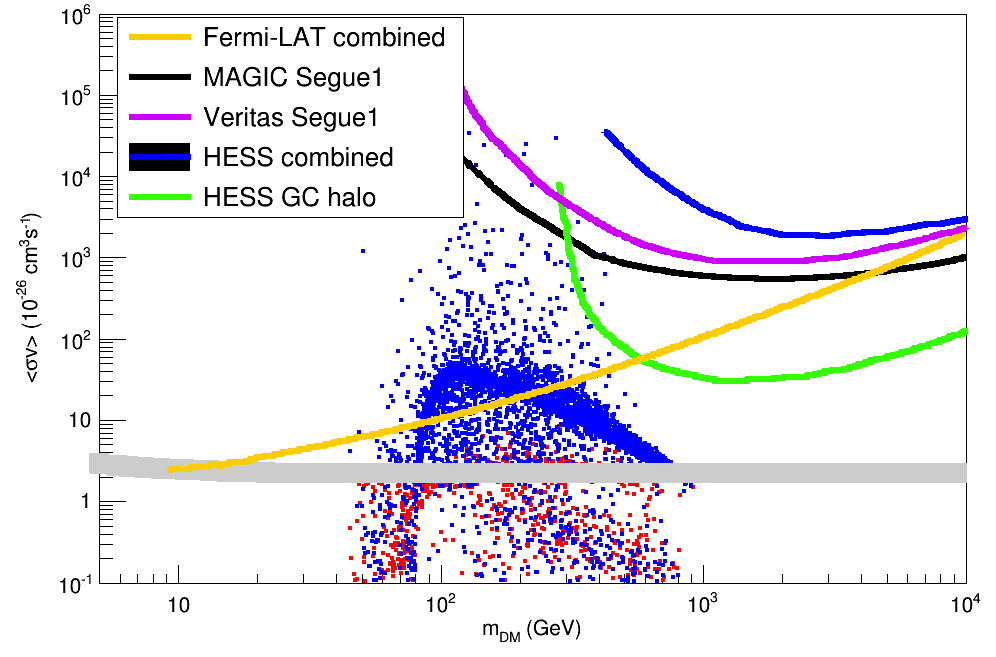}
\includegraphics[width=0.49\columnwidth]{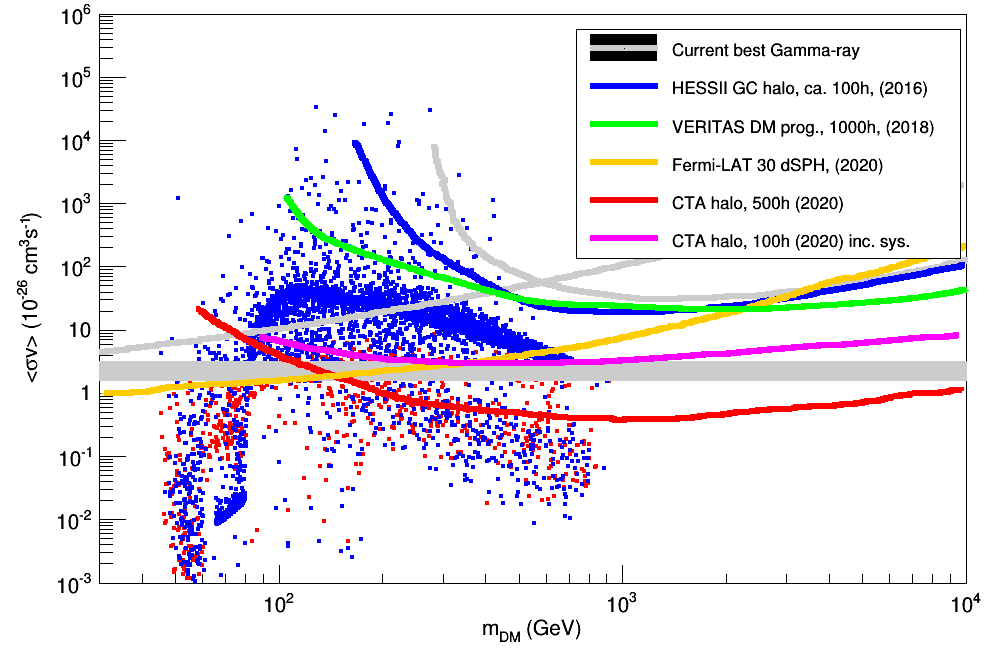}
\caption{Left panel: current most relevant constraints on annihilation cross-section versus mass for quark-channels. Fermi-LAT dwarfs 
\cite{Ackermann:2013yva}, HESS dwarf galaxies \cite{Abramowski:2014tra}, HESS halo \cite{Abramowski:2011hc}, VERITAS Segue1 \cite{Aliu:2012ga}, MAGIC Segue1
\cite{Aleksic:2013xea}. Right panel: Constraints in the next decade: VERITAS 1000h DM programme \cite{Smith:2013tta}, HESSII (my guess), Fermi-LAT dwarf and CTA \cite{Buckley:2013bha} and CTA including systematic uncertainties \cite{Silverwood:2014yza}. Model points, taken from \cite{Abdo:2010ex}, (as in other figures) represent MSSM-7 with red colour: consistent with cosmological dark matter density, blue colour: annihilation cross-section allowed to be smaller. }
\label{fig:results}
\end{center}
\end{figure}

\noindent
For IACTs, the vicinity of the GC constitutes the most promising and relatively robust target providing the most stringent constraints (by more than one dex) above around 1 TeV \cite{Abramowski:2011hc}.  Above a 100 GeV, the systematics of the diffuse gamma-ray background are not important. For cusped profiles, the uncertainty in the obtained constraints is about a factor 2 \cite{Abramowski:2011hc}, for cored profiles the background subtraction technique employed causes a significant decrease of sensitivity which however can be remedied by applying a dedicated pointing technique. The decisive advantage that dwarf galaxies provide (robust density determination and negligible astrophysical background) is not as important for IACTs, nevertheless dwarfs are more robust with respect to assumptions on cored versus cusped density profile. Constraints are therefore presented by MAGIC and VERITAS for deep exposures of the Segue 1 dwarf galaxy \cite{Aleksic:2013xea,Aliu:2012ga}, by HESS of a combined analysis of their dwarf observations \cite{Abramowski:2014tra}. The most important current results for b-quark channels are summarized in Fig. \ref{fig:results}, left panel. In the right panel we show future expectation with existing and future instruments. HESS has augmented the four-telescope array with a fifth telescope in 2012 with lower energy threshold\footnote{I present a guesstimate of the performance of a GC halo analysis, not based on officially approved collaboration statements}. VERITAS envisages a 1000h programme of DM searches with dwarf galaxies \cite{Smith:2013tta}. A Fermi-LAT prediction shown is taken from \cite{Buckley:2013bha}, it does not include an improvement in analysis methods and indeed the recently presented PASS8 results are already excluding WIMPS below the thermal cross-section up to 100 GeV. The next major step forward will be the Cherenkov Telescope Array, CTA, an array of about 80 telescopes with a factor 10 improved sensitivity and an extended energy range from about 10 GeV to 40 TeV \cite{Consortium:2010bc}. A prediction more or less based on scaling performances of IACTs arrived at the conclusion that the thermal WIMP cross-section could be probed at masses between 100 GeV and 10 TeV \cite{Doro:2012xx,Buckley:2013bha} . For CTA, systematics on the cosmic ray background estimate (ratio between off and on region acceptance) and irreducible diffuse emission can become critical in the limit of large event statistics. An attempt to account for both has been presented in \cite{Silverwood:2014yza} with somewhat dimmed expectations. Also pair conversion telescopes are planned, foremost GAMMA-400 \cite{Bongi:2014zwa}, the DArk Matter Particle Explorer (DAMPE) \cite{Dampe:2014} and the High Energy cosmic Radiation Detector (HERD) \cite{Zhang:2014haa}. Both have in common very deep calorimeters, enabling energy resolution of the order of 1 \%. The latter has an envisaged effective area of about twice the Fermi-LAT, and main progress can be expected in the area of spectral feature detection. e.g. \cite{Bergstrom:2012vd}.

\section{Charged cosmic-ray probe: a clear signal -- but of what?}

\noindent
The main signature for DM in charged cosmic rays are in the anti-proton and positron channel. Anti-particles are very rarely produced in secondary processes and even a small addition of anti-particles produced in WIMP annihilation could give rise to a detectable signal, revealing itself as a rise in the positron to electron or antiproton to proton ratio, conveniently the ratio is taken to cancel acceptance systematics which should affect particles and anti-particles similarly. A smoking gun signature is provided by anti-deuteron, as the expected signal is up to 4 dex larger than the background for energies below about 1 GeV/n.

\noindent
The main instruments for positrons/anti-proton are PAMELA \cite{Picozza:2006nm} and more recently AMS \cite{Kounine:2012ega}. PAMELA, launched in 2006, on a Russian satellite is a spectrometer with a geometric factor (GF) of about 20 cm sr, AMS, launched in 2011,  is operating on the International Space Station with a GF which is  larger by 1 to 2 dex, depending on analysis.  Also the maximum rigidity that can be reconstructed is larger by a factor 2 ($\sim$ 2000 GV). \\

\begin{figure}[!ht]
\begin{center}
\includegraphics[width=0.5\columnwidth]{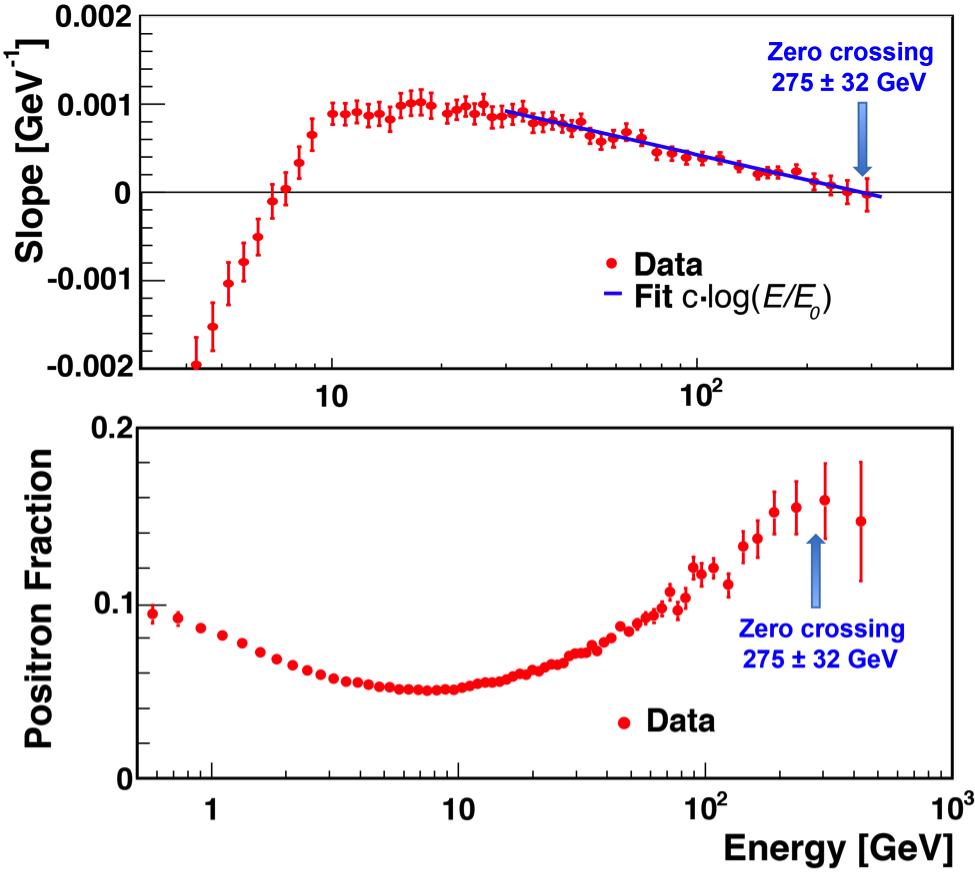}
\caption{Slope parameter and positron fraction as measured recently by AMS \cite{Accardo:2014lma}.}
\label{fig:AMS}
\end{center}
\end{figure}

\noindent
The results of the PAMELA experiment, presented in 2008,  were a game changer: a rising positron fraction \cite{Adriani:2008zr} together with a hard (electron+positron) spectrum presented by the Fermi-LAT \cite{Ackermann:2010ij} strongly suggests an additional source of positrons.  AMS confirmed the rise of the positron fraction with much improved events statistics and extended the measurement in energy \cite{Aguilar:2013qda, Accardo:2014lma}, see Fig. \ref{fig:AMS}, most recently also claiming the detection of the flattening of the positron fraction.  If the positron excess is interpreted as of WIMP origin, generically, the result prefers relatively large particle masses ($\sim$ TeV), preferably annihilation into leptonic states,  and requires some kind of mechanism to enhance the annihilation cross-section (substructure or Sommerfeld enhancement \cite{Bergstrom:2009fa}. Considering the updated AMS results, it is hard to accommodate the hard electron spectrum of Fermi unless the decay proceeds via light mediator states \cite{Cholis:2013psa}. The fact that there is an unexplained excess is of course exciting, but there are well motivated conventional sources that could provide the extra positrons: foremost pulsars have been discussed, either a single close-by mature pulsar such as Monogem or Geminga, e.g. \cite{Yuksel:2008rf,Profumo:2008ms} or a sum of the Milky Way pulsar population, e.g. \cite{Cholis:2013psa}. It has also been suggested that excess positrons could be produced by standard cosmic ray sources, in particular Super Nova Remnants (SNR). It is conjectured that SNRs could produce positrons inside with a harder spectrum than the usual secondary positrons \cite{Blasi:2009hv,Ahlers:2009ae}, a hypothesis testable with future measurements of the proton/anti-proton and B/C ratios \cite{Mertsch:2011gd}. It is likely that the spectral shape of the positron fraction alone will not be sufficient to determine the origin of the positrons \cite{Delahaye:2014osa}. One other approach could then me to measure the direction of the flux, or to be more precise the anisotropy. A dipole anisotropy of the positron+electron rate in the direction of for example Monogem could be detectable with IACTs  \cite{Linden:2013mqa} thanks to their immense effective area, though it is remains to be seen  if the required sensitivity can be reached given the involved systematic uncertainties. \\

\noindent
A so far largely unexplored probe of WIMP annihilation is anti-deuterons. In contrast to positrons, antideuterons provide a smoking-gun signal of WIMPs, as predicted backgrounds of secondaries is several orders of magnitude lower. AMS has anti-deuteron capabalities, but the most sensitve future detector is the General Antiparticle Spectrometer (GAPS), a baloon borne detector that performed a successful prototype flight in 2012 \cite{vonDoetinchem:2013oxa}.

\section{Neutrino-probe: direct detection in disguise}
The neutrino channel can be used for probing WIMP annihilation, targeting pretty much the same sources as gamma rays, i.e. GC, halo, dwarf galaxies as well as galaxy clusters.  The main edge for neutrino telescopes, however, is the search for WIMP induced neutrinos from the Sun. WIMPs are captured in the Sun by scattering and as the Sun is dominated by hydrogen, neutrino searches from the Sun are very sensitive to the spin-dependent part of the WIMP-nucleon cross-section, a quantity usually only accessible by direct detection experiments.\\ 

\noindent
The instruments searching for neutrinos resulting from WIMP annihilation are IceCube \cite{Achterberg:2006md} and ANTARES \cite{Collaboration:2011nsa}. IceCube, situated at the south-pole consists of 86 strings, each equipped with 60 optical modules (photosensors). As strings have been added, analysis has been performed in different configurations. The latest published results are on 79 string configuration (IC79). The standard operation mode is to detect neutrinos in the northern hemisphere, whereas if fiducialization is applied, also the southern hemisphere, in particular the GC comes within reach. ANTARES is considerably smaller. It has 885 strings on 12 strings and is situated in the Mediterranean sea, close to Toulon, France. In standard operation mode (without fiducialization) is to observe the southern hemisphere, in particular the GC becomes visible.\\

\begin{figure}[!ht]
\includegraphics[width=0.49\columnwidth]{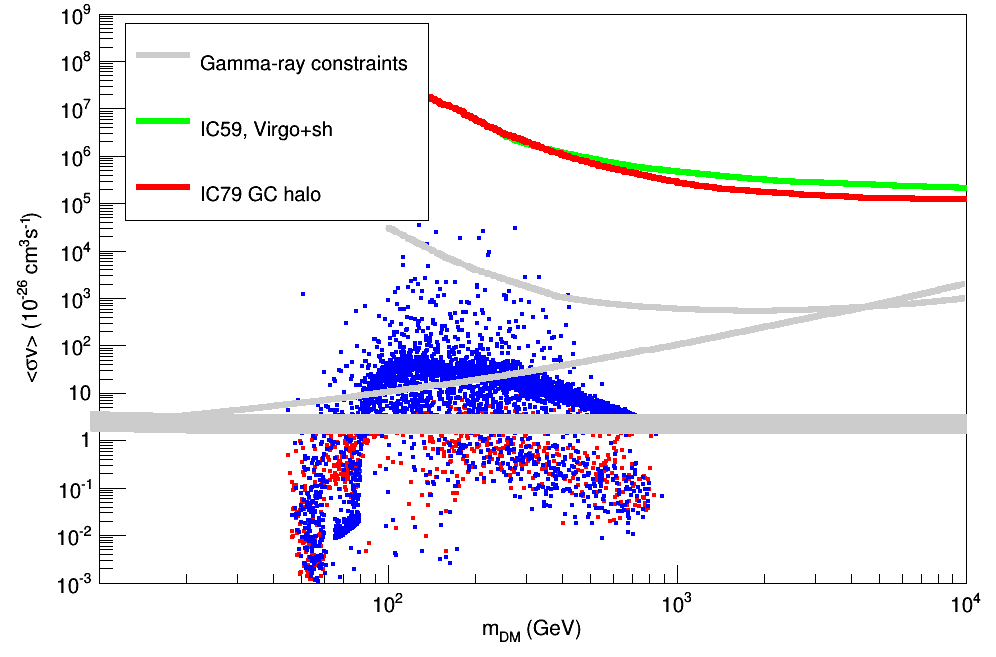}
\includegraphics[width=0.49\columnwidth]{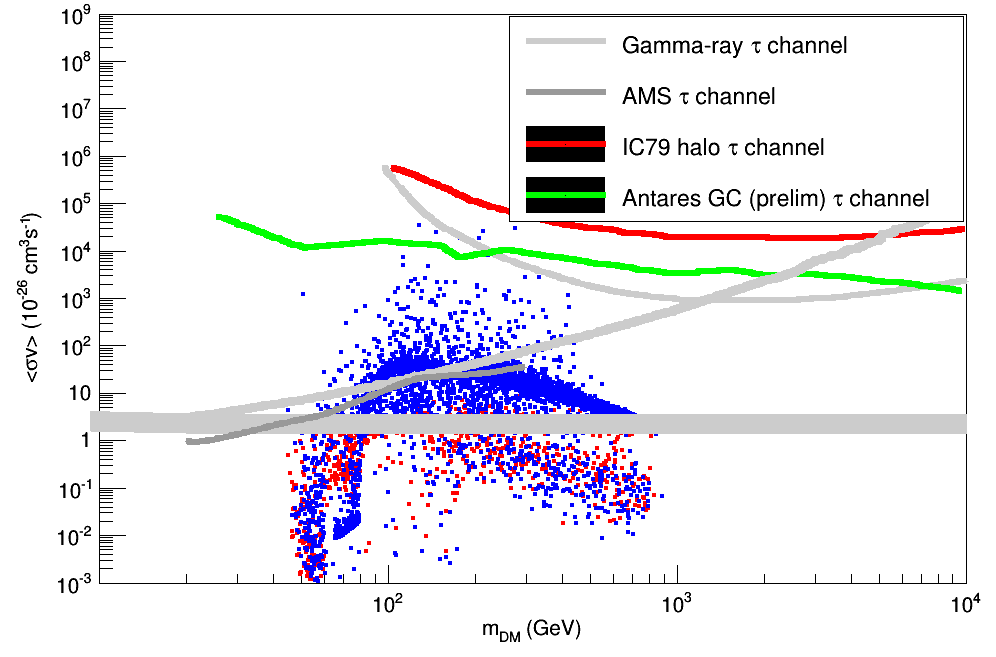}
\caption{Most relevant constraints on annihilation cross section as a function of WIMP mass as obtained by neutrino telescopes. Left panel: IceCube observations of the
Virgo cluster (including substructure boost) (IC59 Virgo+sh)  \cite{Aartsen:2013dxa} and the multipole analysis of the halo, 79 string configuration \cite{Aartsen:2014hva}, gamma-ray constraints are included for comparison, Right panel:  $\tau$-channel, ANTARES GC halo \cite{Bertin:2013mma} and IceCube 79 string halo. Also included corresponding gamma-ray and AMS constraints \cite{Bergstrom:2013jra}.}
\label{fig:neutrinos}
\end{figure}

\noindent
The present status is summarized in figure \ref{fig:neutrinos}. Results have been presented for the Galactic halo \cite{Aartsen:2014hva}, clusters of galaxies and dwarf galaxies by IceCube \cite{Aartsen:2013dxa}  and for the GC halo by ANTARES \cite{Bertin:2013mma} . For annihilation preferably into quarks, upper limits obtained by neutrino telescopes are three to four dex larger than those obtained by IACTS, becoming competitive only for very high WIMP masses close to the unitarity bound. For WIMPs annihilating into the tau-channel, limits are comparable and neutrino telescopes become more sensitive for masses above a few TeV.\\

\begin{figure}[!ht]
\begin{center}
\includegraphics[width=0.5\columnwidth]{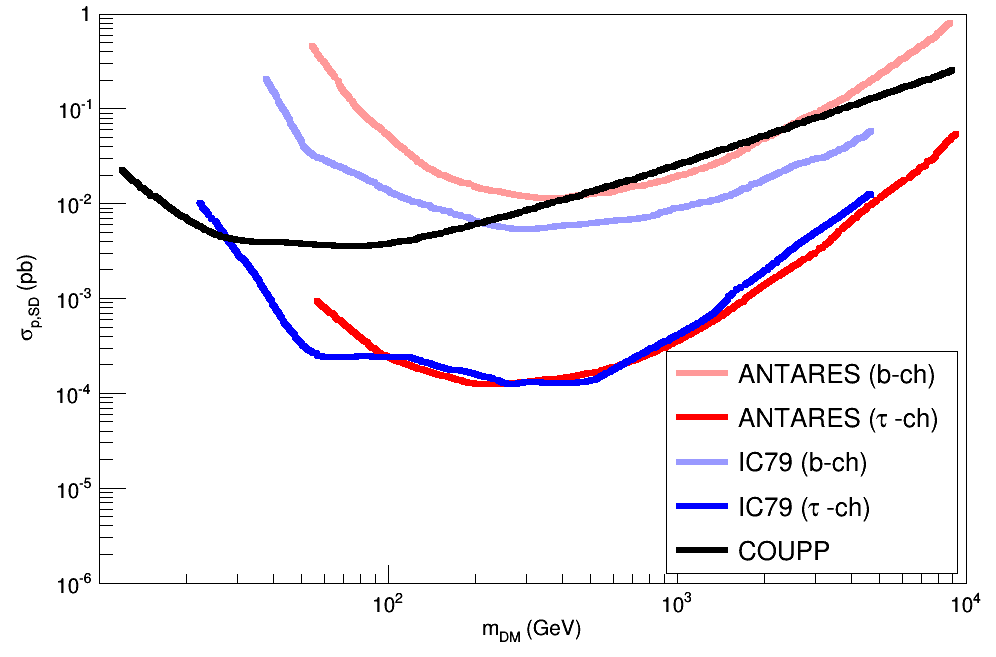}
\caption{Constraints on the spin-dependent part of the WIMP scattering cross-section versus WIMP mass. IC79 \cite{Aartsen:2012kia} and preliminary limits by ANTARES \cite{Bertin:2013mma} are compared to results by COUPP \cite{Behnke:2012ys}.}
\end{center}
\label{fig:dir}
\end{figure}

\noindent
As mentioned previously, neutrino telescopes become more relevant for searches for WIMP induced neutrino emission from the Sun. Here, world-leading constraints on the spin-dependent WIMP nucleon cross-section can be obtained. The annihilation rate is after equilibrium is reached dominated by the capture rate, i.e. by the scattering cross-section. Due to the hydrogen dominance in the Sun, the resulting sensitivity to the spin-dependent part of the scattering cross-section is excellent. The present status is summarized in Fig. \ref{fig:dir}. Also for SD cross-section, the upper limits depend strongly on the assumptions regarding the annihilation channel, leptonic channels being preferred. For masses above about 200 GeV constraints obtained by IceCube \cite{Aartsen:2012kia} are roughly a factor 2 better than by COUPP \cite{Behnke:2012ys} (the most sensitive direct detection experiment), assuming annihilation into b-quarks preferably. Assuming $\tau$-quark annihilation, the constraints obtained by IceCube and ANTARES are up to 2 dex better than COUPP at 200 GeV.

\section{Summary \& Discussion}
With the advent of the Fermi-LAT indirect detection of WIMPs has entered a pivotal period. Indirect detection with gamma-rays is the main work horse providing first time a serious and robust probe of the thermal WIMP annihilation cross-section up to masses of about 100 GeV.  As experiments reach sensitivities that probe the most interesting parameter space, ``discoveries'' appear, for example the 130 GeV line or the Galactic Centre GeV excess. The existence of the latter is confirmed, however the explanation is unclear. The postulated DM  origin of the GC excess is not confirmed by the most recent search for DM induced gamma rays in in dwarf spheroidal galaxies (the PASS 8 iteration), but it is true that ``absence of evidence is not the same as evidence of absence'', as J-factor uncertainties in the Galactic Centre region could alleviate the tension.\\

\noindent 
The rise of the positron fraction observed by PAMELA and confirmed and extended by AMS indicates an extra source of positions. A DM interpretation\footnote{preferring TeV dark matter and leptonic channels, unlike the gamma ray excess which 
generically prefers dark matter masses below 100 GeV.} is possible, but increasingly difficult. A close-by or a superposition of a number of pulsars, as well as secondary production in SNR (soon to be tested)  may provide a viable explanation. The neutrino probe for the annihilation cross-section only for very large masses (beyond 10 TeV) and preferably leptonic channels. Neutrinos have an edge when they probe an observable normally connected with direct detection, namely the spin-dependent part of the nucleon-WIMP scattering cross-section. Here neutrino telescopes provide world leading limits by searching for WIMP induced neutrinos from the Sun.


\bigskip
\section{Acknowledgments}
Support of the  Swedish National Space Board, the Swedish Resarch Council and the Knut and Alice Wallenberg, foundation is acknowledged. JC is Wallenberg Academy Fellow. Christian Farnier is thanked for providing the MSSM-7 model points.

%
%

%
%
%
%
 
\end{document}